\documentclass{ieeeaccess}

\usepackage{cite}
\usepackage{amsmath,amssymb,amsfonts}
\usepackage{algorithmic}
\usepackage{graphicx}
\usepackage{textcomp}

\usepackage{caption}
\usepackage{epsfig, amsmath, amssymb, cite, booktabs, array, multirow}
\usepackage{dblfloatfix} 
\usepackage{arydshln}
\usepackage{mathtools}
\usepackage{mathrsfs}
\usepackage{boldline}

\usepackage{subcaption}

\def\BibTeX{{\rm B\kern-.05em{\sc i\kern-.025em b}\kern-.08em
    T\kern-.1667em\lower.7ex\hbox{E}\kern-.125emX}}

\begin{document}
\history{Date of publication xxxx 00, 0000, date of current version xxxx 00, 0000.}
\doi{10.1109/ACCESS.2019.DOI}

\title{Disentangled speaker and nuisance attribute embedding for robust speaker verification}
\author{\uppercase{Woo Hyun Kang}\authorrefmark{1}, \IEEEmembership{Student Member, IEEE},
	\uppercase{Sung Hwan Mun}\authorrefmark{2}, \IEEEmembership{Student Member, IEEE},
	\uppercase{Min Hyun Han}\authorrefmark{3}, \IEEEmembership{Student Member, IEEE}, 
and \uppercase{Nam Soo Kim}\authorrefmark{4}, \IEEEmembership{Senior Member, IEEE}}
\address[1]{Department of Electrical and Computer Engineering and INMC,	Seoul National University, 1 Gwanak-ro, Gwanak-gu, Seoul 08826, Korea (e-mail: whkang@hi.snu.ac.kr)}
\address[2]{Department of Electrical and Computer Engineering and INMC,	Seoul National University, 1 Gwanak-ro, Gwanak-gu, Seoul 08826, Korea (e-mail: shmun@hi.snu.ac.kr)}
\address[3]{Department of Electrical and Computer Engineering and INMC,	Seoul National University, 1 Gwanak-ro, Gwanak-gu, Seoul 08826, Korea (e-mail: mhhan@hi.snu.ac.kr)}
\address[4]{Department of Electrical and Computer Engineering and INMC,	Seoul National University, 1 Gwanak-ro, Gwanak-gu, Seoul 08826, Korea (e-mail: nkim@snu.ac.kr)}
\tfootnote{This research was supported by Projects for Research and Development of Police science and Technology under Center for Research and Development of Police science and Technology and Korean National Police Agency funded by the Ministry of Science, ICT and Future Planning (PA-J000001-2017-101).}

\markboth
{Kang \headeretal: Disentangled speaker and nuisance attribute embedding for robust speaker verification}
{Kang \headeretal: Disentangled speaker and nuisance attribute embedding for robust speaker verification}

\corresp{Corresponding author: Nam Soo Kim (e-mail: nkim@snu.ac.kr).}

\begin{abstract}
Over the recent years, various deep learning-based embedding methods have been proposed and have shown impressive performance in speaker verification.
However, as in most of the classical embedding techniques, the deep learning-based methods are known to suffer from severe performance degradation when dealing with speech samples with different conditions (e.g., recording devices, emotional states).
In this paper, we propose a novel fully supervised training method for extracting a speaker embedding vector disentangled from the variability caused by the nuisance attributes.
The proposed framework was compared with the conventional deep learning-based embedding methods using the RSR2015 and VoxCeleb1 dataset.
Experimental results show that the proposed approach can extract speaker embeddings robust to channel and emotional variability.
\end{abstract}

\begin{keywords}
speech embedding, speaker verification, domain disentanglement, deep learning.
\end{keywords}

\titlepgskip=-15pt

\maketitle

\section{Introduction} \label{intro}
\label{sec:introduction}
\PARstart{S}{peaker} verification is the task of verifying the claimed speaker identity based on the given speech samples and has become a key technology for personal authentication in many commercial applications, forensics and law enforcement \cite{spkmag}.
Commonly, an utterance-level fixed-dimensional vectors (i.e. embedding vectors) are extracted from the enrollment and test speech samples and then fed into a scoring algorithm (e.g., cosine distance, probabilistic linear discriminant analysis) to measure their similarity or likelihood of being spoken by the same speaker.
Over the past years, the i-vector framework has been one of the most dominant approaches for speech embedding \cite{ivec}, \cite{ivec_2}.
The widespread popularity of the i-vector framework in the speaker verification community can be attributed to its ability to summarize the distributive pattern of the speech with a relatively small amount of training data in an unsupervised manner.

In recent years, various methods have been proposed utilizing deep learning architectures for extracting embedding vectors and have shown better performance than the i-vector framework when a large amount of training data is available \cite{xvec}.
In \cite{dvec}, a deep neural network (DNN) for frame-level speaker identification was trained and the averaged activation from the last hidden layer, namely, the d-vector, was taken as the embedding vector for text-dependent speaker verification.
In \cite{xvec, xvec2}, a speaker identification model consisting of a frame-level network and a segment-level network was trained and the hidden layer activation of the segment-level network (i.e. x-vector) was extracted as the embedding vector.
In \cite{dvec_e2e}, long short-term memory (LSTM) layers were adopted to capture the contextual information within the d-vector, and the embedding network was trained to directly optimize the verification score (e.g., cosine similarity) in an end-to-end fashion.
The end-to-end d-vector framework was further enhanced in \cite{dvec_e2e_att} by applying different weight (i.e. attention) to each frame-level activation while obtaining the d-vector, which enables the embedding network to attend more on the frames with relatively higher amount of speaker-dependent information.
In \cite{dvec_ge2e}, a generalized end-to-end loss function, which optimizes the embedding vector to move towards the centroid of the true speaker while departing away from the centroid of the most confusing speaker, was introduced to train the end-to-end d-vector system more efficiently.
In \cite{spk_vae} and \cite{spk_ali}, a variational autoencoder (VAE)-based architecture was trained in an unsupervised manner to extract an embedding vector for short-duration speaker verification.
Despite their success in well-matched conditions, the deep learning-based embedding methods are vulnerable to the performance degradation caused by mismatched conditions (e.g., channel, noise) \cite{grl1}.

In real life applications, numerous factors can contribute to the mismatches in speaker verification \cite{spkmag}.
Especially in forensic situations, channel mismatch often occurs since police officers usually acquire voice recordings using various recording devices (e.g., hidden microphones, mobile phones) \cite{forensic}.
Such variation in recording devices is known to cause variability to the speech distribution, which leads to low speaker identification or verification performance.

Recently, many attempts have been made to extract an embedding vector robust to mismatched conditions.
Conventionally, various researches focused on adapting the back-end scoring model (e.g., PLDA) \cite{vaeda} or training the embedding network with an augmented dataset containing various nuisance variability \cite{cycleda}.
These methods are proven to be effective when the dataset for the target condition (e.g., noisy evaluation domain) is scarce, but since these methods do not intervene during the embedding extraction, their performance may be bottlenecked by the speaker discriminative capability of the embedding network.
Unlike the aforementioned domain adaptation techniques, there have been several methods which aim to directly disentangle the undesired variability while extracting the speaker embeddings.
In \cite{grl1, grl2}, inspired by the usage of gradient reversal strategy in image classification \cite{im_grl}, \cite{im_grl2} and robust speech recognition \cite{asr_grl1, asr_grl2}, the embedding networks were trained to minimize the speaker classification error while maximizing the error of the subtask (e.g., noise or channel type classification) with the use of gradient reversal layer.
Although the gradient reversal strategy has shown meaningful improvement in performance, domain adversarial training using gradient reversal layer is known to be very unstable and sensitive to hyper-parameter setting \cite{grl_bad}.
In \cite{antiloss}, the embedding network was trained to maximize the error of a subtask (i.e. noise type classification) by using an adversarial training strategy similarly to the generative adversarial network (GAN) \cite{gan}.
The speaker embedding network and the noise classification network are trained competitively; the noise classification network is trained to discriminate the noise type correctly, and at the same time the embedding network is trained to discriminate the speaker while having high uncertainty on the noise type.
When training the speaker embedding network, bit-inverted one-hot labels (i.e. anti-labels) were used for noise classification, which would force the embedding network to output a wrong noise label equally.
Though the anti-label strategy has proven its strength in noise-robust speaker embedding \cite{antiloss}, adversarial training is known to be extremely unstable and difficult \cite{gan_bad}.

In this paper, we propose a novel approach to disentangle the nuisance attribute information from the speaker embedding vector without the use of gradient reversal or adversarial training.
The proposed method employs an embedding network similar to the conventional methods (e.g., d-vector and x-vector).  
However, unlike the conventional embedding networks, which produce a single embedding vector per utterance, the proposed embedding network simultaneously extracts a speaker- and nuisance attribute-dependent (e.g., recording device-, emotion-dependent) embedding vectors, hence we call the proposed technique joint factor embedding (JFE).
In the JFE technique, the embedding network is trained in a fully supervised manner simultaneously with the speaker and nuisance attribute (e.g., channel, emotion) discriminator networks where each discriminator is trained to take the embedding vector as input and identify their respective targets.
Analogous to the conventional speaker embedding systems, the proposed embedding network is trained to produce a speaker embedding vector with high speaker discriminability.
On the other hand, to disentangle the non-speaker information from the speaker embedding vector, we propose two different ways to increase the nuisance attribute uncertainty inherent in the speaker embedding vector.
One way is to train the embedding network to extract a speaker embedding vector to maximize the entropy in nuisance attribute identification, and the other is to decrease the relevancy between the speaker and nuisance embedding vectors by minimizing the mean absolute Pearson's correlation (MAPC) \cite{mapc}.

In order to evaluate the performance of the proposed system in a realistic scenario, we conducted a set of experiments using two datasets:
\begin{itemize}
	\item RSR2015 Part 3 dataset: a random digits strings speaker verification corpus consisting of speech samples recorded from 6 different hand-held devices \cite{rsr2015}, \cite{rsr2015_2}.
	\item VoxCeleb1 dataset: a text-independent speaker verification corpus consisting of speech samples with 8 different emotional states \cite{vox1}.
\end{itemize}
The experimental results show that the proposed method outperforms the conventional disentanglement methods (i.e. gradient reversal, anti-label) in terms of equal error rate (EER).
Moreover, the proposed system performed better than the conventional x-vector on short duration speech samples, which is likely to lack significant phonetic information.

The contributions of this paper are as follows:
\begin{itemize}
	\item We propose a new method to train a speaker embedding network robust to nuisance attributes, which can be done easily without the use of adversarial training or gradient reversal learning.
	\item We compared the proposed speaker embedding technique with conventional methods for multi-device and emotional speaker verification.
	\item We experimented the proposed speaker embedding technique on speech utterances with various durations.
\end{itemize}

The rest of this paper is organized as follows: We first briefly describe the conventional embedding network architecture and disentanglement methods based on gradient reversal and anti-label in Section \ref{d-vector}.
In Section \ref{j-vector}, the newly proposed JFE scheme is presented.
The experiments and results are shown in Section \ref{experiment}.
Finally, Section \ref{conclusion} concludes the paper.

\section{Deep learning-based speaker embedding} \label{d-vector}

\subsection{Deep embedding network}

\begin{figure}
	\begin{subfigure}[b]{0.5\textwidth}
		\centering
		\includegraphics[width=0.95\linewidth]{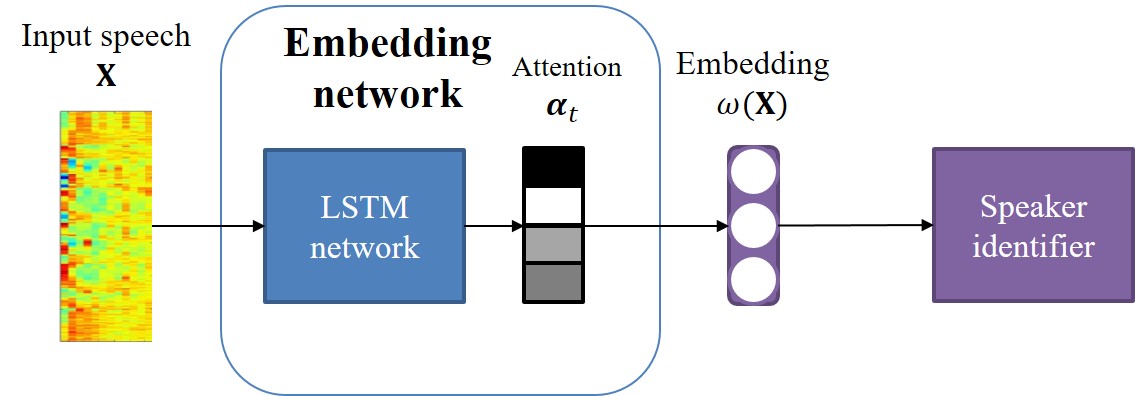}
		\caption{}
		\label{dvec_softmax_struct}
	\end{subfigure}
	\begin{subfigure}[b]{0.5\textwidth}
		\centering
		\includegraphics[width=0.95\linewidth]{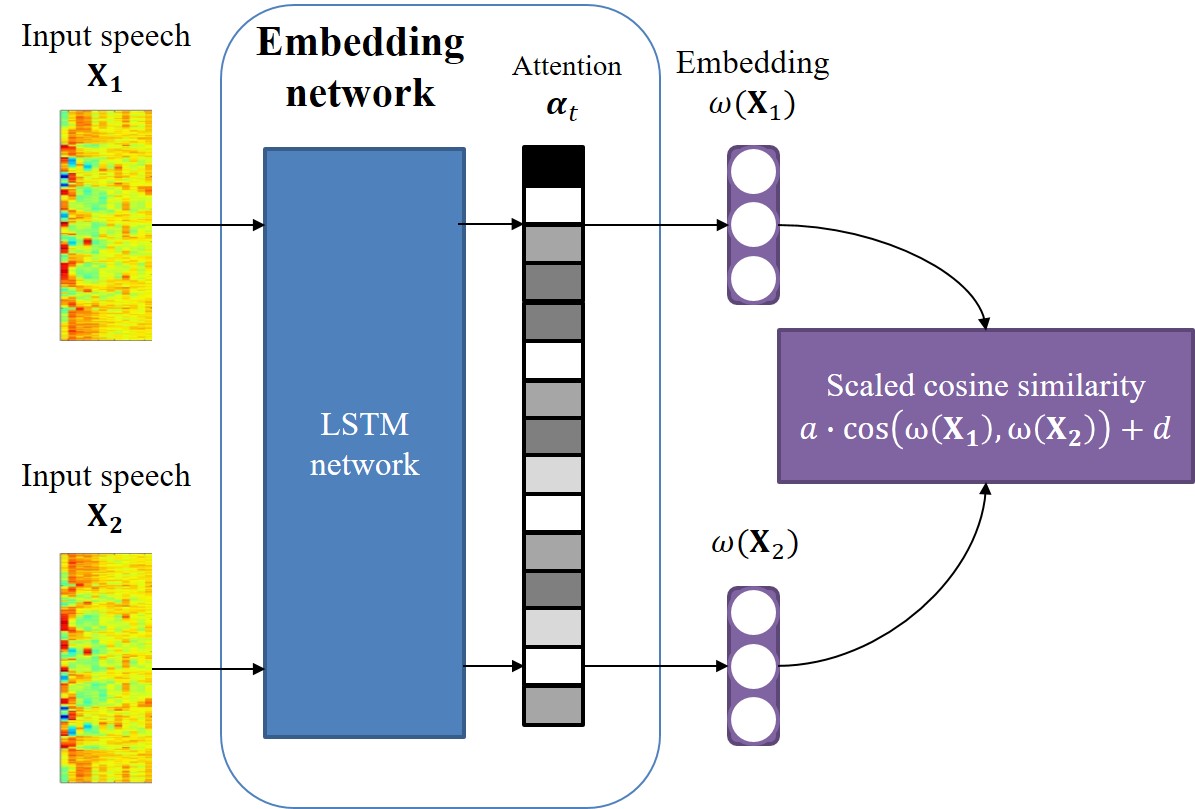}
		\caption{}
		\label{dvec_e2e_struct}
	\end{subfigure}
	\caption{(a) LSTM-based d-vector system trained with softmax loss. (b) LSTM-based d-vector system trained with end-to-end loss.}
\end{figure}

Two of the most widely used speaker embedding techniques are the LSTM-based d-vector \cite{dvec_ge2e} and the TDNN (time-delay DNN)-based x-vector system \cite{xvec}.
In both frameworks, given a speech utterance $\mathbf{X}$ with $T$ frames, a sequence of frame-level acoustic features $\{\mathbf{x}_{1},...,\mathbf{x}_{T}\}$ extracted from $\mathbf{X}$ is fed into the frame-level network.
In the d-vector system, one of most widely used technique for text-dependent speaker recognition, the frame-level network is composed of LSTM layers, which helps capture the temporal correlation.
On the other hand, the frame-level network of the x-vector system consists of TDNN layers, which is often used for text-independent speaker recognition.
Once the frame-level outputs $\{\mathbf{h}_{1},...,\mathbf{h}_{T}\}$ are obtained, they are aggregated to obtain an utterance-level representation.
One way of aggregating the frame-level outputs is to compute the weighted average as
\begin{equation} \label{attention_1}
\mathbf{\omega}=\sum_{t=1}^{T}{\alpha}_{t}\mathbf{h}_{t}
\end{equation}
where $\alpha_{t}{\in}[0, 1]$ is a normalized weight, which is computed by
\begin{equation} \label{attention_2}
{\alpha}_{t}=\frac{{\exp}(e_t)}{\sum_{t=1}^{T}{\exp}(e_t)}.
\end{equation}
In (\ref{attention_2}), the frame-level score (i.e. attention) $e_t$ is computed as follows:
\begin{equation} \label{attention_3}
e_t=\mathbf{v}^{\intercal}_{t}{\tanh}(\mathbf{W}_{t}\mathbf{h}_{t}+\mathbf{b}_{t})
\end{equation}
where $\mathbf{v}_t$, $\mathbf{W}_t$, and $\mathbf{b}_t$ are trainable parameters and superscript $\intercal$ indicates transpose operation.
By using different weight for each frame, speech frames with relatively higher speaker-relevancy can contribute more to the embedding vector.

The embedding network is trained by either minimizing the speaker identification loss \cite{dvec} or directly optimizing the verification performance (i.e. end-to-end speaker verification) \cite{dvec_ge2e}.
In the first case (i.e. embedding network trained for identification), as shown in Fig. \ref{dvec_softmax_struct}, a feed-forward neural network for classifying the speakers in the training set is trained jointly with the embedding network.
The speaker classification network takes the utterance-level representation $\mathbf{\omega}$ as input and has an $N$-dimensional softmax output $\tilde{\mathbf{y}}(\mathbf{\omega})$ where $N$ corresponds to the number of training speakers.
Given the one-hot speaker label $\mathbf{y}$, the embedding and classification networks are trained to minimize the following cross-entropy loss function:
\begin{equation} \label{softmax_dvec}
\mathbf{L}_{spkr}=-\sum_{n=1}^{N}{\mathbf{y}_{n}}{\log}{\tilde{\mathbf{y}}_{n}(\mathbf{\omega})}
\end{equation}
where $\mathbf{y}_{n}$ and $\tilde{\mathbf{y}}_{n}(\mathbf{\omega})$ are the $n^{th}$ components of $\mathbf{y}$ and $\tilde{\mathbf{y}}(\mathbf{\omega})$, respectively.

For training the end-to-end speaker verification system (i.e. embedding network trained for verification), a mini-batch of $J{\times}K$ utterances is fed into the embedding network where the mini-batch is composed of $J$ speakers, and each speaker has $K$ utterances.
As depicted in Fig. \ref{dvec_e2e_struct}, the scaled cosine similarity between each embedding vector and the centroid of the embedding vectors from each speaker are computed by
\begin{equation} \label{e2e_dvec_1}
\mathbf{S}_{jk,i}=a{\cdot}{\cos}(\mathbf{\omega}_{jk}, \mathbf{c}_{i})+d
\end{equation}
where $a$ and $d$ are trainable parameters, and ${\cos}(\mathbf{\omega}_{jk}, \mathbf{c}_{i})$ is the cosine similarity between the utterance-level representation extracted from the $k^{th}$ utterance of the $j^{th}$ speaker $\mathbf{\omega}_{jk}$ and the centroid of the $i^{th}$ speaker's utterance-level representations $\mathbf{c}_{i}$ ($1~{\leq}~j, i~{\leq}~J$ and $1~{\leq}~k~{\leq}~K$).
For each utterance-level representation $\mathbf{\omega}_{jk}$ in the mini-batch, the embedding network is trained to maximize the following end-to-end loss function:
\begin{equation} \label{e2e_dvec_2}
\mathbf{L}_{e2e}=\mathbf{S}_{jk,j}-{\log}\sum_{i=1,i{\neq}j}^{J}{\exp}(\mathbf{S}_{jk,i}).
\end{equation}
The end-to-end system is known to outperform the softmax method when a large amount of dataset is used for training \cite{xvec2}, \cite{dvec_e2e}.

Once the embedding network is trained, the utterance-level representation $\mathbf{\omega}$ \cite{dvec_ge2e}, or the hidden layer activation of the speaker classification network \cite{xvec} can be used as the speaker embedding vector.

\subsection{Conventional disentanglement methods}

\begin{figure}
	\begin{subfigure}[b]{0.5\textwidth}
		\centering
		\includegraphics[width=0.95\linewidth]{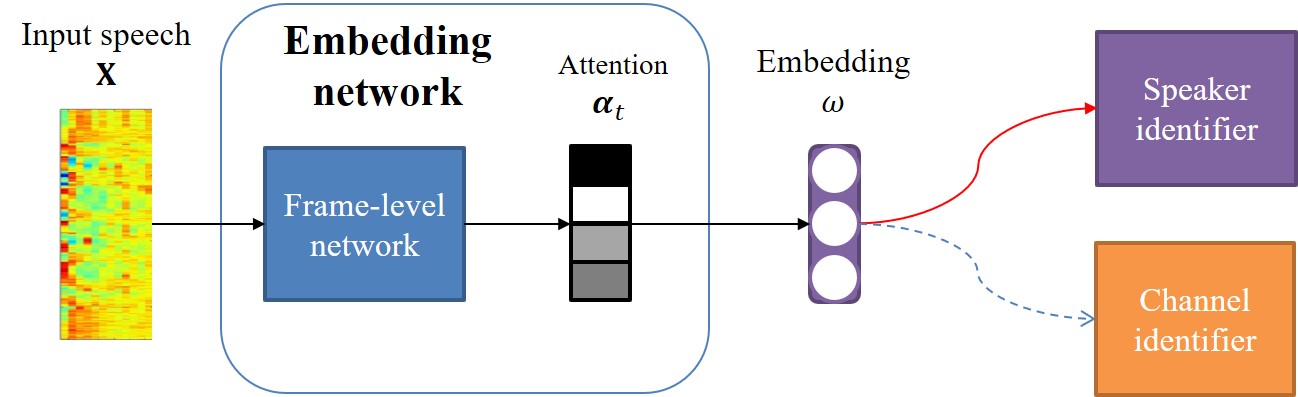}
		\caption{}
		\label{mtl_struct}
	\end{subfigure}
	\begin{subfigure}[b]{0.5\textwidth}
		\centering
		\includegraphics[width=0.95\linewidth]{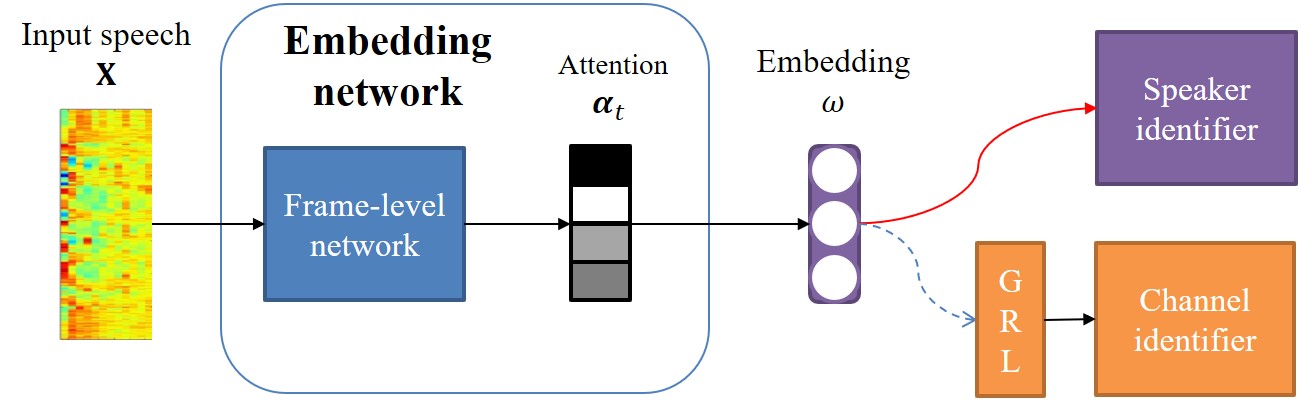}
		\caption{}
		\label{grl_struct}
	\end{subfigure}
	\caption{(a) Standard multi-task learning (MTL) architecture. (b) Domain adversarial training via gradient reversal layer (GRL).}
\end{figure}

Recently, disentangling various non-speaker factors (e.g., channel type, noise type, noise-level) from the embedding vector has become an important issue in speaker verification \cite{grl1, grl2, antiloss}.
Most of the techniques developed to address this issue are based on the multi-task learning (MTL) approaches \cite{mtl} where the embedding network is trained to optimize in two tasks: main task (i.e. speaker classification) and subtask (e.g., channel classification) as shown in Fig. \ref{mtl_struct}.
The objective of the MTL-based disentanglement technique is to achieve the best performance in the main task while degrading the performance in the subtask.

\subsubsection{Gradient reversal strategy}
One way to achieve this is the gradient reversal strategy, which has shown meaningful performance in channel-robust \cite{grl2} and noise-robust \cite{grl1} speaker verification.
As shown in Fig. \ref{grl_struct}, the gradient reversal strategy adds a gradient reversal layer (GRL) \cite{im_grl} between the subtask network and the embedding network.
Let ${\theta}_{emb}$, ${\theta}_{main}$, ${\theta}_{sub}$ denote the parameters for the embedding, main task, and subtask networks.
The GRL performs identity transformation on the input during forward propagation and reverses the gradient by multiplying a negative scalar $-\lambda$ during backpropagation.
When jointly training the networks, the parameters are updated as
\begin{equation} \label{grl_1}
\theta_{emb}~{\leftarrow}~\theta_{emb}-l\cdot(\frac{{\partial}{\mathbf{L}_{main}}}{{\partial}{\theta_{emb}}}-\lambda\frac{{\partial}{\mathbf{L}_{sub}}}{{\partial}{\theta_{emb}}}),
\end{equation}
\begin{equation} \label{grl_2}
\theta_{main}~{\leftarrow}~\theta_{main}-l\cdot(\frac{{\partial}{\mathbf{L}_{main}}}{{\partial}{\theta_{main}}}),
\end{equation}
\begin{equation} \label{grl_3}
\theta_{sub}~{\leftarrow}~\theta_{sub}-l\cdot(\frac{{\partial}{\mathbf{L}_{sub}}}{{\partial}{\theta_{sub}}})
\end{equation}
where $l$, $\mathbf{L}_{main}$, and $\mathbf{L}_{sub}$ are the learning rate, loss functions for the main task and subtask, respectively.
For extracting a channel-robust embedding for speaker verification, $\mathbf{L}_{main}$ would be the speaker cross-entropy $\mathbf{L}_{spkr}$ defined in (\ref{softmax_dvec}), and $\mathbf{L}_{sub}$ would be the channel cross-entropy which can be computed as follows:
\begin{equation} \label{softmax_device}
\mathbf{L}_{chan}=-\sum_{m=1}^{M}{\mathbf{r}_{m}}{\log}{\tilde{\mathbf{r}}_{m}(\mathbf{\omega})}
\end{equation}
where $M$ is the number of different channels (e.g., recording devices) in the training set, ${\mathbf{r}_{m}}$ and $\tilde{\mathbf{r}}_{m}(\mathbf{\omega})$ are the $m^{th}$ component of the one-hot channel label ${\mathbf{r}}$ and channel classifier's softmax output $\tilde{\mathbf{r}}(\mathbf{\omega})$, respectively.

\subsubsection{Anti-loss strategy}
Another way to achieve disentanglement is by training the embedding network and the subtask network in a competitive manner via adversarial training \cite{antiloss}.
The subtask network is trained to classify the channel identity correctly given the embedding vector as in (\ref{softmax_device}).
On the other hand, the main task and embedding networks are trained to discriminate the speaker by minimizing (\ref{softmax_dvec}) but not to perform well on the subtask.
In order to ensure high uncertainty on the subtask, \cite{antiloss} introduces anti-label when computing the cross-entropy for the subtask.
The anti-label is obtained by flipping each bit in the one-hot label vector.
This indicates that for channel disentanglement, the anti-loss can be computed as follows:
\begin{equation} \label{antiloss_device}
\mathbf{L}_{anti-dev}=-\sum_{m=1}^{M}(1-{\mathbf{r}_{m}}){\log}{\tilde{\mathbf{r}}_{m}(\mathbf{\omega})}.
\end{equation}
By minimizing $\mathbf{L}_{anti-dev}$ and $\mathbf{L}_{speaker}$ simultaneously, the embedding network would be trained to produce a speaker discriminative embedding vector which is robust to channel variability.

\section{Joint factor embedding} \label{j-vector}

\subsection{Joint factor embedding network architecture}

\begin{figure*}
	\centering
	\includegraphics[width=0.8\linewidth]{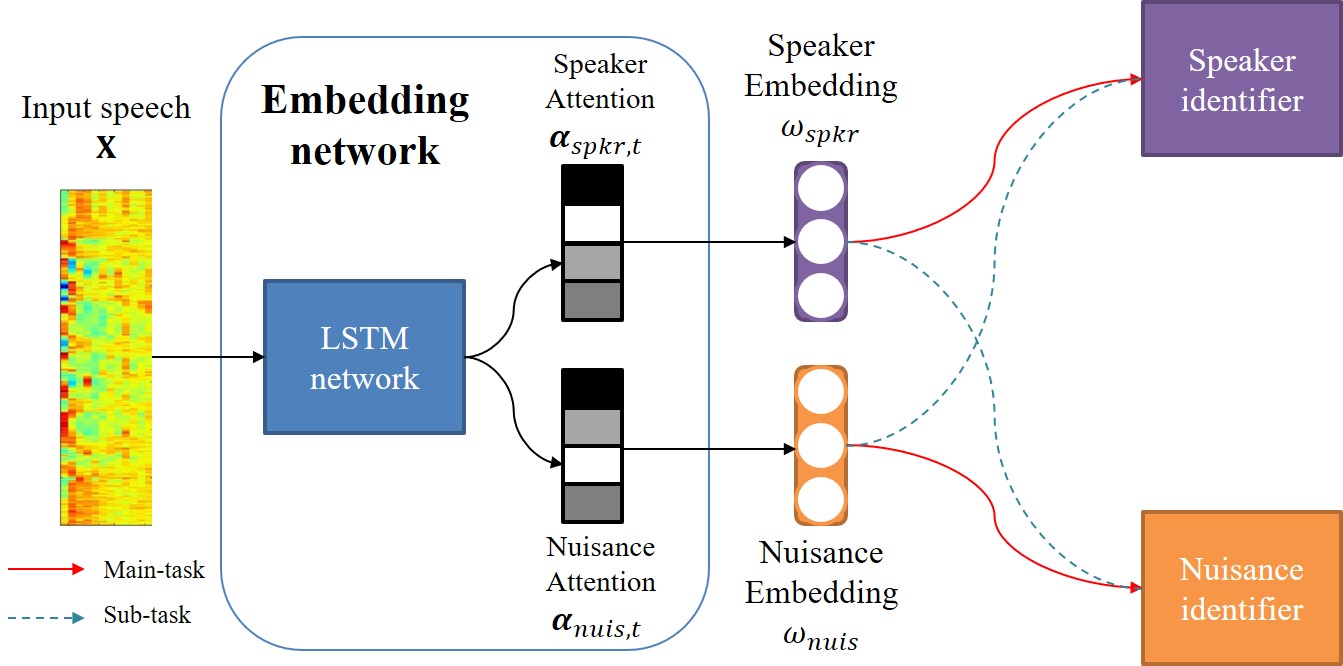}
	\caption{The architecture of the proposed joint factor embedding system.}
	\label{jvec_struct}
\end{figure*}

Analogous to the conventional disentanglement techniques \cite{grl1, grl2, antiloss}, the proposed method is based on the MTL framework.
However, as depicted in Fig. \ref{jvec_struct}, unlike the standard MTL embedding system, the embedding network of the proposed framework extracts two different embedding vectors simultaneously: speaker embedding ${\omega}_{spkr}$ and nuisance embedding ${\omega_{nuis}}$.
The speaker embedding vector ${\omega_{spkr}}$ is trained to be dependent solely on the speaker variability while the nuisance embedding vector ${\omega_{nuis}}$ is trained to be dependent on the nuisance (e.g., channel, emotion) variability only.
When obtaining ${\omega_{spkr}}$ and ${\omega_{nuis}}$, different weights are used for aggregating the frame-level outputs as
\begin{equation} \label{speaker_attention}
\mathbf{\omega}_{spkr}=\sum_{t=1}^{T}{\alpha}_{spkr,t}\mathbf{h}_{t},
\end{equation}
\begin{equation} \label{device_attention}
\mathbf{\omega}_{nuis}=\sum_{t=1}^{T}{\alpha}_{nuis,t}\mathbf{h}_{t}
\end{equation}
where ${\alpha}_{spkr,t}$ and ${\alpha}_{nuis,t}$ are the speaker and nuisance weights for attention, respectively, which are obtained as in (\ref{attention_2}).
The reason why we use separate attention weights for obtaining ${\omega}_{spkr}$ and ${\omega_{nuis}}$ is that we assume that frames with high speaker-dependent information are not always guaranteed to have high nuisance attribute-dependent information.
For instance, speaker-dependent information will be high on speech frames, while channel-dependent information will be rather consistent across all frames since even non-speech frames are affected by the recording channel.
Once the embedding vectors are extracted, both ${\omega_{spkr}}$ and ${\omega_{nuis}}$ are fed into the speaker and nuisance classification networks.

\subsection{Training for joint factor embedding}
\begin{table}[b]
	\renewcommand{\arraystretch}{1.3}
	\caption{Main tasks and subtasks for the embedding vectors of the joint factor embedding scheme.}
	\label{task_jvec}
	\centering
	\begin{tabular}{c|c|c}
		\hline
		&  Main task & Subtask \\
		\hline
		\hline
		
		$\omega_{spkr}$   &   Speaker classification & Nuisance classification\\
		\hline
		
		$\omega_{nuis}$    &   Nuisance classification & Speaker classification\\
		\hline
	\end{tabular}
\end{table}

\subsubsection{Discriminative training}
As described in Table \ref{task_jvec}, the embedding vectors ${\omega_{spkr}}$ and ${\omega_{nuis}}$ are trained with different main task and subtask specifications.
In order to maximize the discriminability on their main tasks, the following cross-entropy loss functions are minimized:
\begin{equation} \label{spkr_spkr_jvec}
\mathbf{L}_{s-s,CE}=-\sum_{n=1}^{N}{\mathbf{y}_{n}}{\log}{\tilde{\mathbf{y}}_{n}(\mathbf{\omega}_{spkr})},
\end{equation}
\begin{equation} \label{dev_dev_jvec}
\mathbf{L}_{c-c,CE}=-\sum_{m=1}^{M}{\mathbf{r}_{m}}{\log}{\tilde{\mathbf{r}}_{m}(\mathbf{\omega}_{nuis})}.
\end{equation}
By minimizing (\ref{spkr_spkr_jvec}) and (\ref{dev_dev_jvec}) simultaneously, the embedding network is trained to produce $\omega_{spkr}$ with high speaker-dependent information and $\omega_{nuis}$ with high nuisance attribute-dependent information.
Moreover, the attention weights ${\alpha}_{spkr,t}$ and ${\alpha}_{nuis,t}$ will be trained to focus on the frames with more meaningful information on their main tasks.

\subsubsection{Disentanglement training}
In this paper, we propose two types of loss functions to perform disentanglement in the subtasks of the embedding vectors ${\omega_{spkr}}$ and ${\omega_{nuis}}$.
One way for disentanglement is to directly maximize the entropy (or uncertainty) on their subtasks while training.
For ${\omega_{spkr}}$ and ${\omega_{nuis}}$, the entropies \cite{entropy} on their subtasks can be computed as
\begin{equation} \label{spkr_dev_jvec}
\mathbf{L}_{s-c,E}=-\sum_{n=1}^{N}{\tilde{\mathbf{y}}_{n}(\mathbf{\omega}_{nuis})}{\log}{\tilde{\mathbf{y}}_{n}(\mathbf{\omega}_{nuis})},
\end{equation}
\begin{equation} \label{dev_spkr_jvec}
\mathbf{L}_{c-s,E}=-\sum_{m=1}^{M}{\tilde{\mathbf{r}}_{m}(\mathbf{\omega}_{spkr})}{\log}{\tilde{\mathbf{r}}_{m}(\mathbf{\omega}_{spkr})}.
\end{equation}
By maximizing (\ref{spkr_dev_jvec}) and (\ref{dev_spkr_jvec}), the uncertainty of the outputs in the subtasks will be maximized, leading the conditional distribution of the subtask classes to approach uniform.

Another way to perform disentanglement is to regularize the embedding vectors ${\omega_{spkr}}$ and ${\omega_{nuis}}$ so as to have low correlation instead of directly maximizing the uncertainty on their subtasks.
This can be achieved by maximizing the negative MAPC \cite{mapc}, which can be computed across the mini-batch by
\begin{equation} \label{mapc}
\mathbf{L}_{nMAPC}=-{\frac{1}{F}}\sum_{f=1}^{F}\frac{|cov(\omega_{spkr, f}, \omega_{nuis, f})|}{std(\omega_{spkr, f})std(\omega_{nuis, f})}
\end{equation}
where $cov$ is the covariance, $std$ is the standard deviation, and $F$, $\omega_{spkr, f}$, $\omega_{nuis, f}$ are the dimensionality of the embedding vectors, $f^{th}$ element of $\omega_{spkr}$ and $\omega_{nuis}$, respectively.
Since zero correlation indicates that the two variables are not related, by minimizing the MAPC between ${\omega_{spkr}}$ and ${\omega_{nuis}}$, the relevancy between the two embedding vectors can be reduced.

The proposed JFE system is trained by simultaneously minimizing the discriminative losses (i.e. cross-entropy) depicted in (\ref{spkr_spkr_jvec}) and (\ref{dev_dev_jvec}), while maximizing the disentanglement loss in (\ref{spkr_dev_jvec}), (\ref{dev_spkr_jvec}), (\ref{mapc}).
In short, the embedding network is trained to minimize the following loss function:
\begin{equation}
\begin{aligned}
\mathbf{L}_{JFE}=
&\mathbf{L}_{s-s,CE}+\mathbf{L}_{c-c,CE} \\
&-\mathbf{L}_{s-c,E}-\mathbf{L}_{c-s,E}-\mathbf{L}_{nMAPC}.
\end{aligned}
\end{equation}
By optimizing the JFE network, the speaker embedding vector ${\omega_{spkr}}$ is trained to be speaker discriminative while having high uncertainty on the nuisance attribute, and the nuisance embedding vector ${\omega_{nuis}}$ aims to be nuisance attribute discriminative while having high uncertainty on the speaker.

\section{Experiments} \label{experiment}

\subsection{Channel disentanglement experiments}

\subsubsection{Database}
In order to evaluate the performance of the proposed technique for a real-life application of speaker verification where multiple recording devices are involved for enrollment and testing, a set of experiments were conducted based on the RSR2015 dataset \cite{rsr2015}, \cite{rsr2015_2}, which is a speaker verification dataset recorded using 6 different hand-held devices (i.e. 1 Samsung Nexus, 2 Samsung Galaxy S, 1 HTC Desire, 1 Samsung Tab, 1 HTC Legend).
For training the embedding networks, we used the \textit{background} and \textit{development} subsets of the RSR2015 dataset Part 3, consisting of utterances (recorded from all six devices) spoken by 194 speakers (100 male and 94 female speakers).

The evaluation was performed according to the RSR2015 Part 3 (random digits string) protocol \cite{rsr_protocol} where 106 speakers (57 male and 49 female speakers) are involved.
From the RSR2015 Part 3 evaluation dataset, the 10-digits strings of sessions 1, 4, 7 were used for enrollment and the 5-digits strings of sessions 2, 3, 5, 6, 8, 9 were used for testing.

\subsubsection{Experimental Setup}

To investigate the effects of the proposed JFE strategy on different embedding architecture, two types of frameworks were used for embedding extraction: d-vector and x-vector.
For the d-vector-based systems, a single 512-dimensional unidirectional LSTM layer with a projection layer \cite{lstm} (projected to 256-dimension) was used.
By aggregating the LSTM outputs via a weighted average as described in (\ref{attention_1}), 256-dimensional embedding vectors were obtained.
Each classification networks (i.e. speaker and channel identifier) consisted of a single 256-dimensional rectified linear unit (ReLU) hidden layer and a softmax output layer where the output size corresponds to the number of speakers or devices within the training set (e.g., 194-dimensional softmax output for speaker classifier and 6-dimensional softmax output for channel classifier).
The acoustic features used in the d-vector-based systems were 19-dimensional Mel-frequency cepstral coefficients (MFCCs) and the log-energy extracted at every 10 ms, using a 20 ms Hamming window.
Together with the delta and delta-delta of the 19-dimensional MFCCs and the log-energy, the frame-level feature used in our experiments was a 60-dimensional vector.

For the x-vector-based systems, 5 TDNN layers were used as the frame-level network as in the Kaldi x-vector recipe \cite{xvec}.
The frame-level output of the last TDNN layer were aggregated via attention pooling (\ref{attention_1}) and followed by a ReLU layer, resulting in a 512-dimensional embedding vector.
The classification networks in the x-vector-based systems consisted of a single 512-dimensional rectified linear unit (ReLU) hidden layer and a softmax output layer.
The acoustic features used in the x-vector-based systems were 30-dimensional MFCCs extracted at every 10 ms, using a 20 ms Hamming window.

The implementation of the embedding systems was done via Tensorflow \cite{tensorflow} and trained using the ADAM optimization technique \cite{adam} with $\beta_1=0.9$ and $\beta_2=0.999$.
All the experimented networks were trained with learning rate 0.001 and batch size 32 for 12,000 iterations.
Cosine similarity was used for computing the verification scores in the experiments.

In our experiments, EER was evaluated as the performance measure.
The EER indicates the error when the false alarm rate (FAR) and the false reject rate (FRR) are the same.

\subsubsection{Comparison between different disentanglement loss terms}

\begin{figure}
	\centering
	\includegraphics[width=\linewidth]{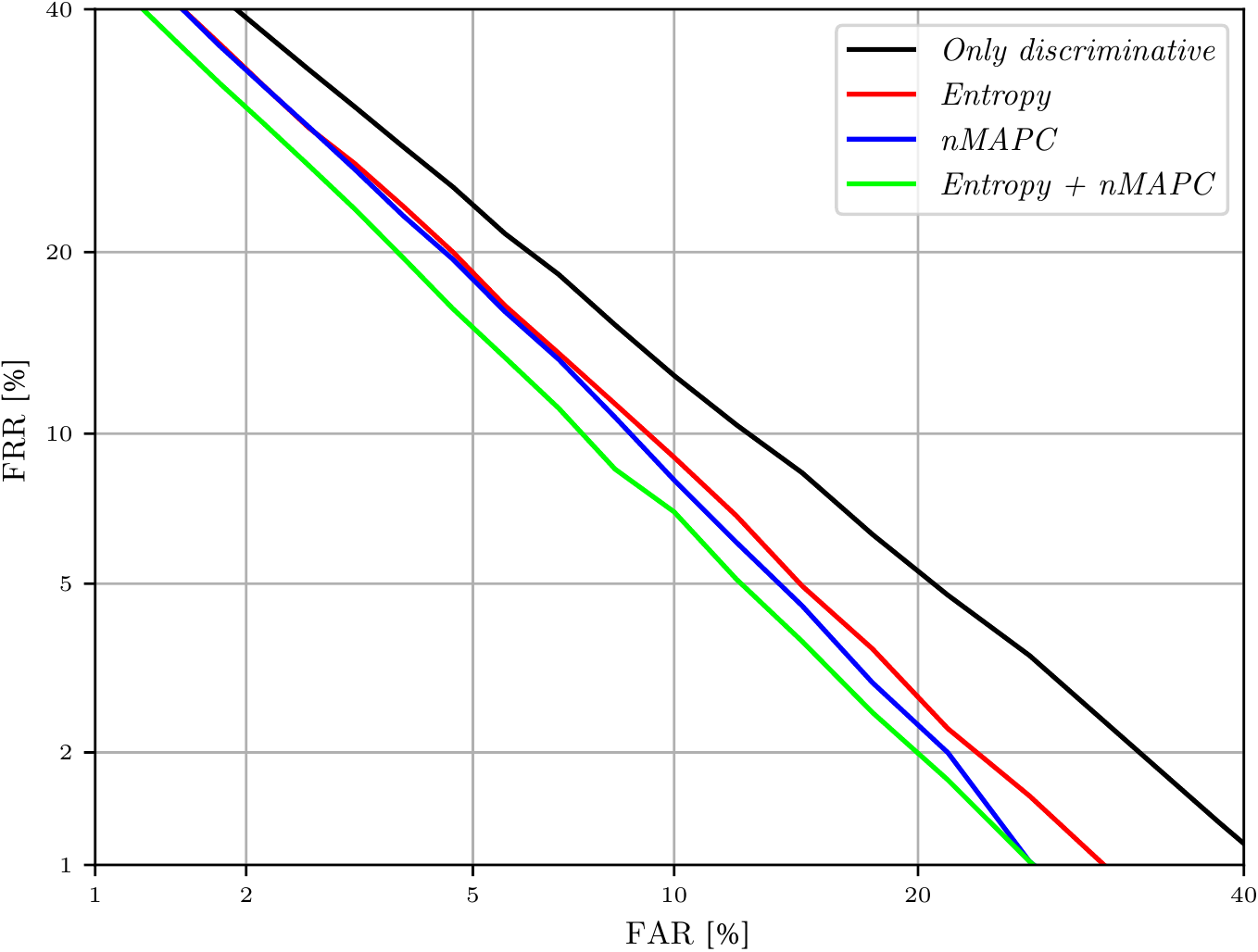}
	\caption{DET curves of the JFE systems trained with various disentanglement losses.}
	\label{loss_det_part3}
\end{figure}

\begin{table}[b]
	\renewcommand{\arraystretch}{1.3}
	\caption{EER (\%) comparison between the speaker embedding vectors extracted from the joint factor embedding networks trained with various disentanglement losses.}
	\label{experiments_loss}
	\centering
	\begin{tabular}{c|c}
		\hline
		 Loss       & EER {[}\%{]}            \\ \hline
		\textit{Only discriminative}  & 11.28                \\ \hline
		\textit{Entropy}             & 9.61                 \\
		\textit{nMAPC}               & 9.25                 \\ \hline
		\textit{Entropy + nMAPC}     & \textbf{8.43}        \\ \hline
	\end{tabular}
\end{table}

In this experiment, we compare the performance of the speaker embeddings obtained from the d-vector-based JFE system trained with different disentanglement loss terms discussed in Section \ref{j-vector}.
The experimented methods are as follows:
\begin{itemize}
	\item {\it{Only discriminative}}: speaker embedding vector extracted from the JFE network trained only with the discriminative loss functions in (\ref{spkr_spkr_jvec}) and (\ref{dev_dev_jvec}) (which is essentially a multi-task learning for the embedding network to encode speaker and nuisance discriminative information),
	\item {\it{Entropy}}: speaker embedding vector extracted from the JFE network trained with the discriminative loss functions in (\ref{spkr_spkr_jvec}), (\ref{dev_dev_jvec}) and the entropy-based disentanglement losses in (\ref{spkr_dev_jvec}) and (\ref{dev_spkr_jvec}),
	\item {\it{nMAPC}}: speaker embedding vector extracted from the JFE network trained with the discriminative loss functions in (\ref{spkr_spkr_jvec}), (\ref{dev_dev_jvec}) and the negative MAPC-based disentanglement losses in (\ref{mapc}),
	\item {\it{Entropy + nMAPC}}: speaker embedding vector extracted from the JFE network trained with the discriminative loss functions in (\ref{spkr_spkr_jvec}), (\ref{dev_dev_jvec}) and both the entropy-based and the negative MAPC-based disentanglement losses in (\ref{spkr_dev_jvec}), (\ref{dev_spkr_jvec}) and (\ref{mapc}).
\end{itemize}

Table \ref{experiments_loss} gives the EER results obtained by using these embeddings.
As shown in the results, the embedding extracted from the JFE networks trained with either \textit{Entropy} or \textit{nMAPC} for disentanglement greatly improved the performance compared to \textit{Only discriminative}, which is essentially a standard MTL embedding technique.
This implies that both \textit{nMAPC} and \textit{Entropy} are capable of training the embedding network to produce speaker embedding vectors disentangled from non-speaker factors.
Especially the \textit{nMAPC} showed relative improvement of 17.99\% compared to \textit{Only discriminative}.
The best verification performance was achieved by using both disentanglement loss terms (i.e. \textit{Entropy + nMAPC}), yielding a relative improvement of 25.27\% in terms of EER.
From this, we could assume that \textit{nMAPC} and \textit{Entropy} are useful for disentangling the channel variability from the speaker embedding.
The DET curves are depicted in Figure \ref{loss_det_part3}.

\subsubsection{Training Analysis}

\begin{figure}
	\centering
	\includegraphics[width=\linewidth]{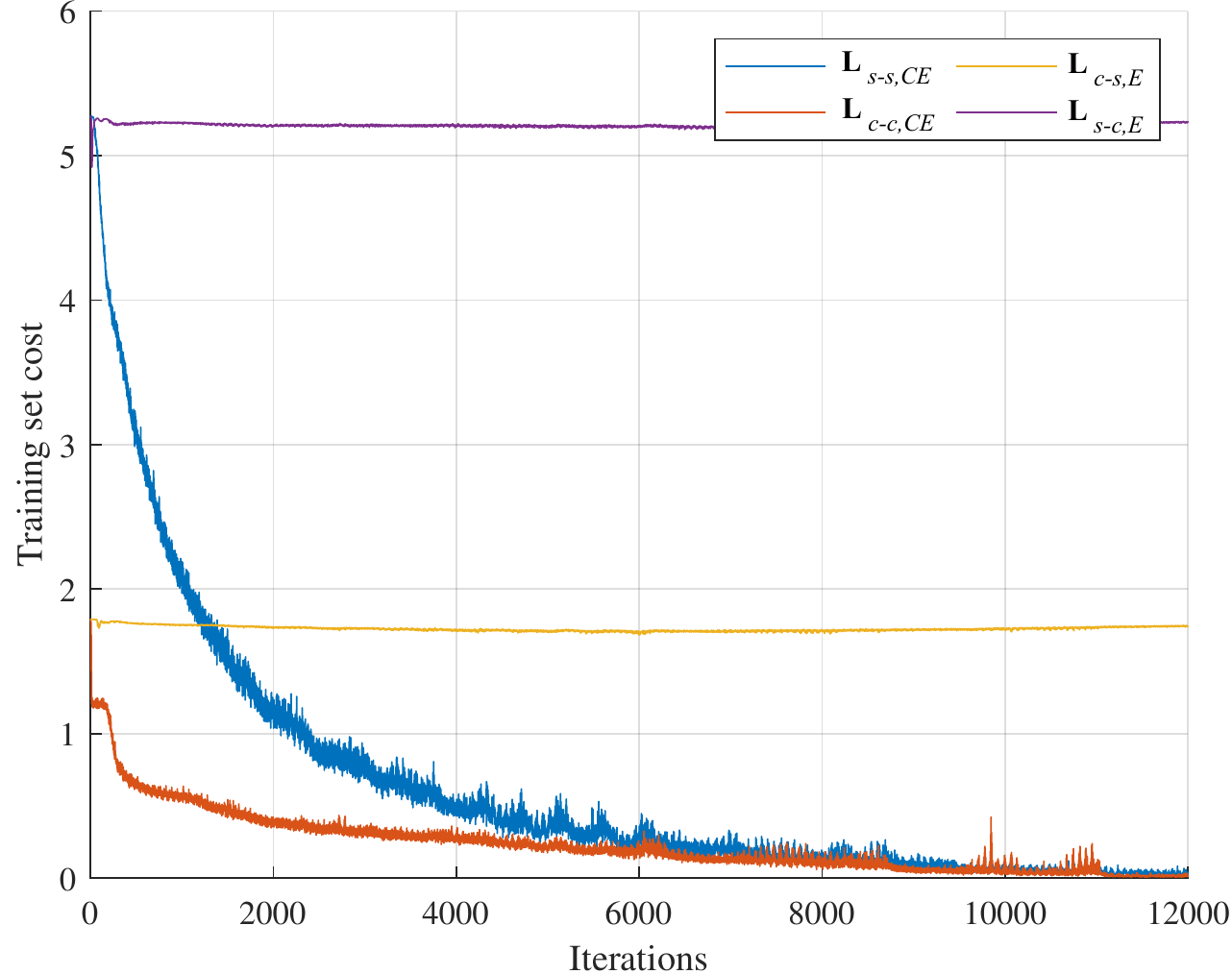}
	\caption{The joint factor embedding training loss values on each iteration.}
	\label{train_cost}
\end{figure}

\begin{figure}
	\begin{subfigure}[b]{0.2\textwidth}
		\centering
		\includegraphics[width=\linewidth]{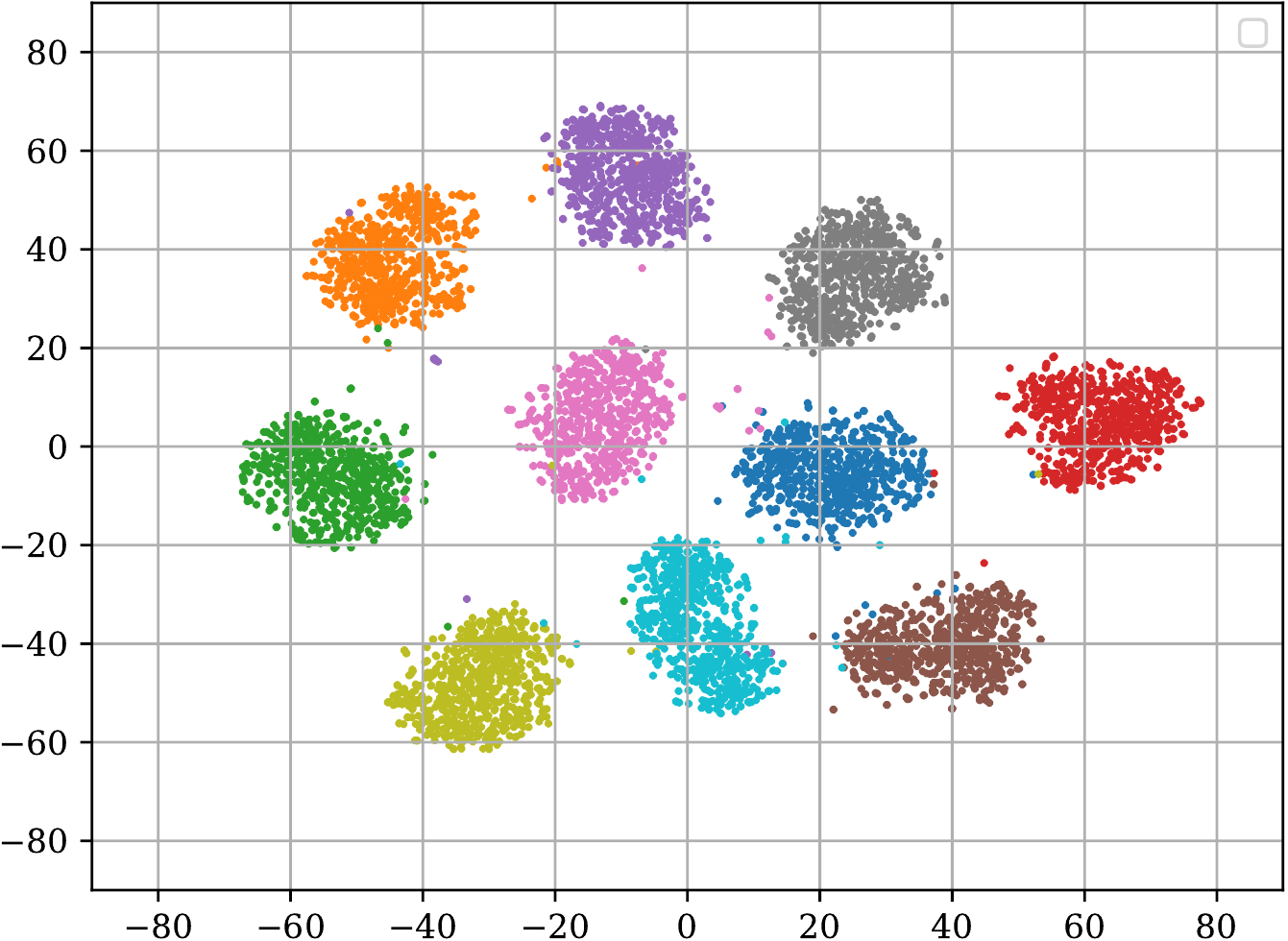}
		\caption{}
		\label{speaker_plot_speaker_emb}
	\end{subfigure}
	\hspace{0.5cm}
	\begin{subfigure}[b]{0.2\textwidth}
		\centering
		\includegraphics[width=\linewidth]{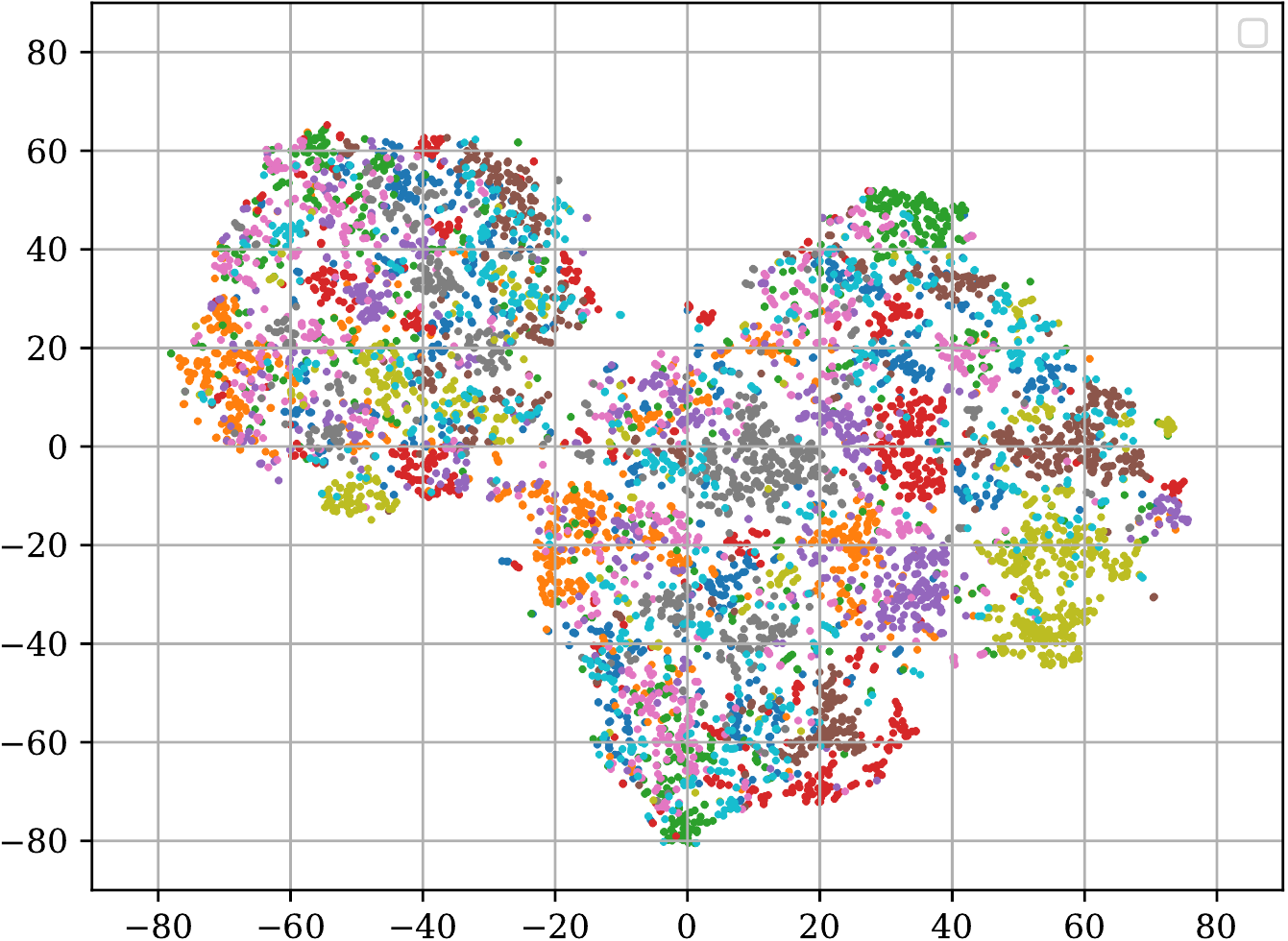}
		\caption{}
		\label{speaker_plot_device_emb}
	\end{subfigure}
	\newline \newline
	\begin{subfigure}[b]{0.2\textwidth}
		\centering
		\includegraphics[width=\linewidth]{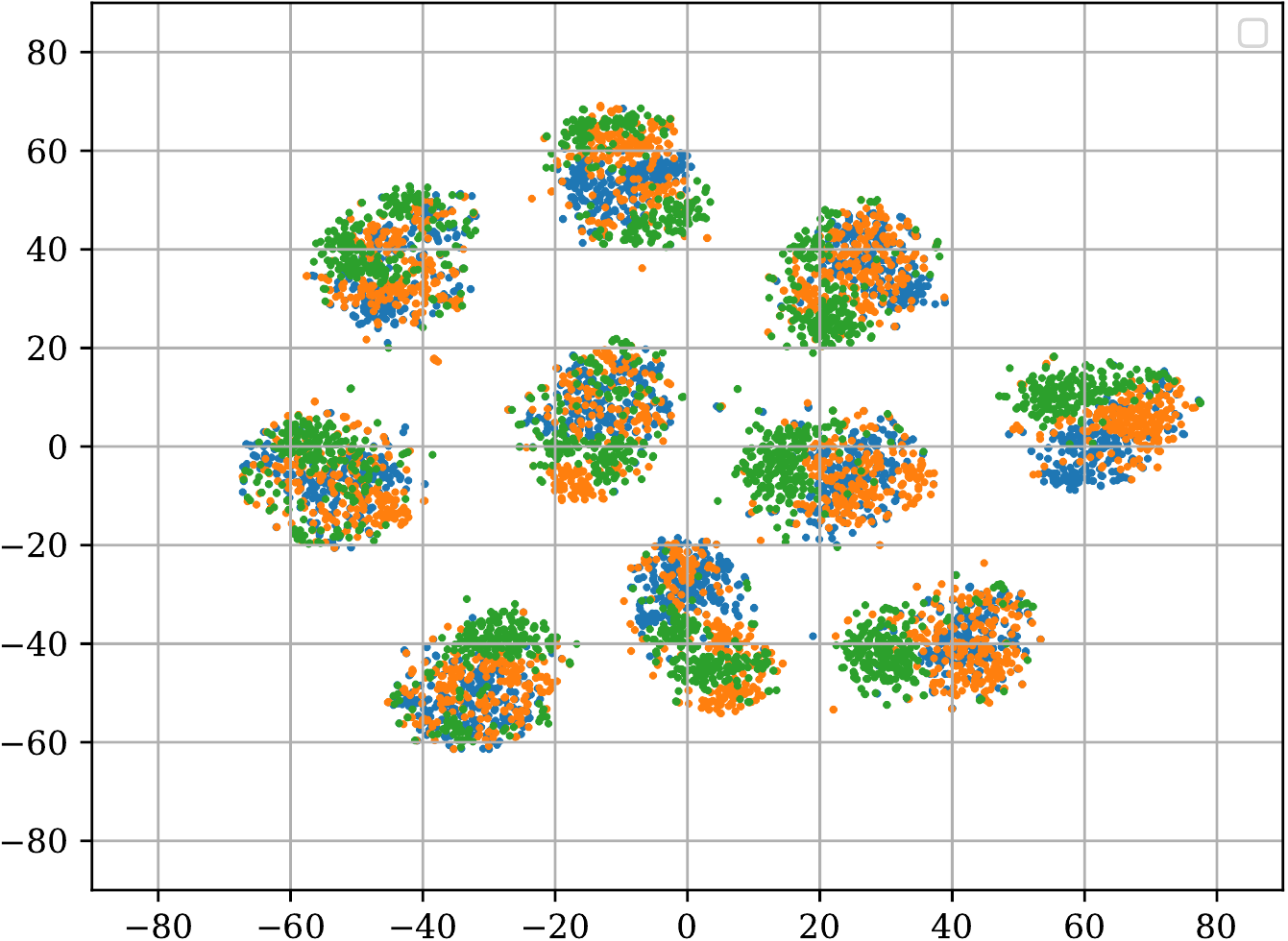}
		\caption{}
		\label{device_plot_speaker_emb}
	\end{subfigure}
	\hspace{0.5cm}
	\begin{subfigure}[b]{0.2\textwidth}
		\centering
		\includegraphics[width=\linewidth]{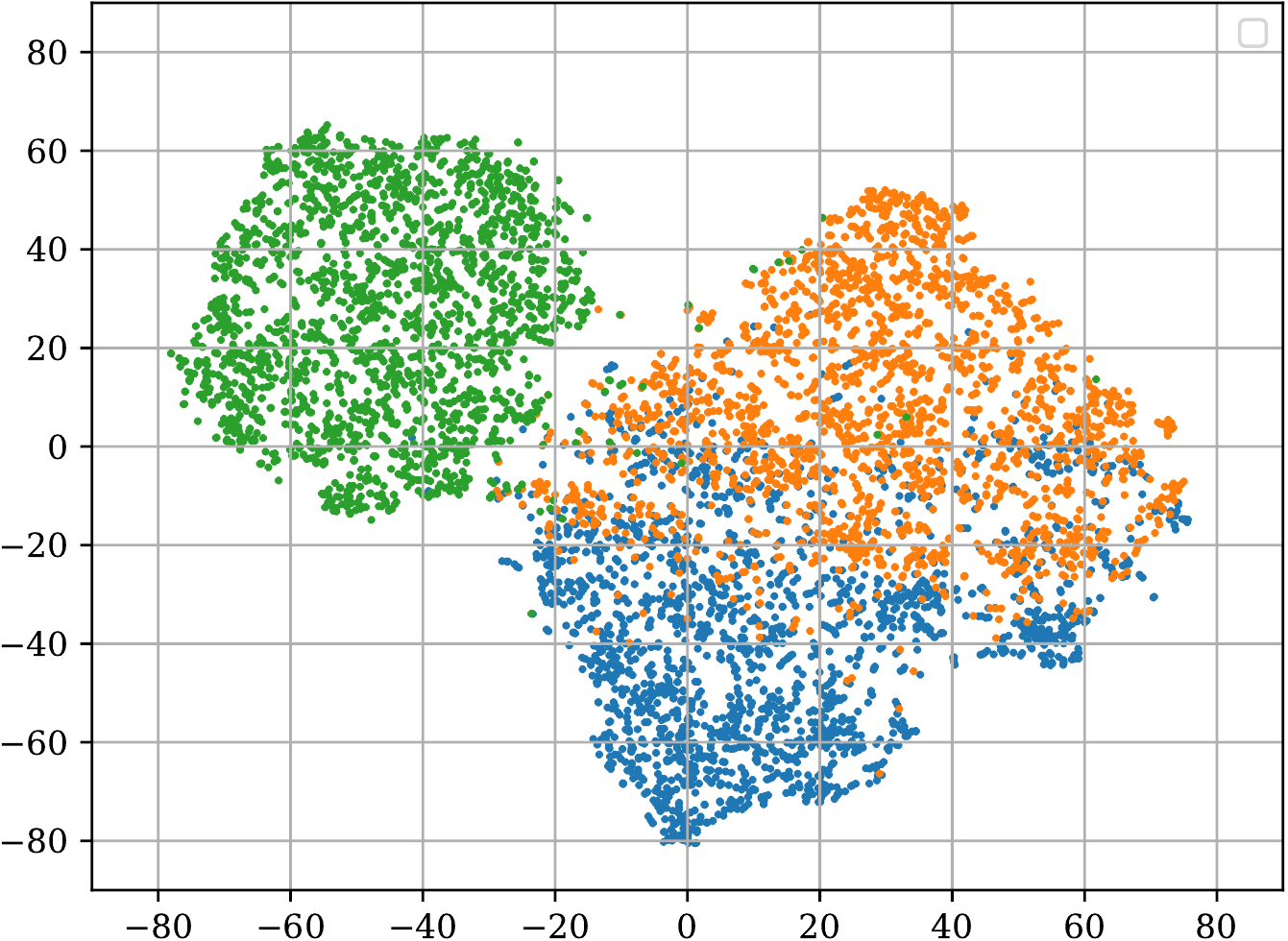}
		\caption{}
		\label{device_plot_device_emb}
	\end{subfigure}
	\caption{t-SNE plot of the speaker and channel embedding vectors extracted from 10 speakers and 3 devices. The x and y axis indicates the $1^{st}$ and $2^{nd}$ dimension of the 2D T-SNE projection, respectively. (a) and (c) are the t-SNE plots of the speaker embedding vectors, and (b) and (d) are the t-SNE plots of the channel embedding vectors. Different colors in (a) and (b) indicate different speakers, and different colors in (c) and (d) indicates different devices.}
	\label{tsne_jvec}
\end{figure}

\begin{figure}
	\begin{subfigure}[b]{0.2\textwidth}
		\centering
		\includegraphics[width=\linewidth]{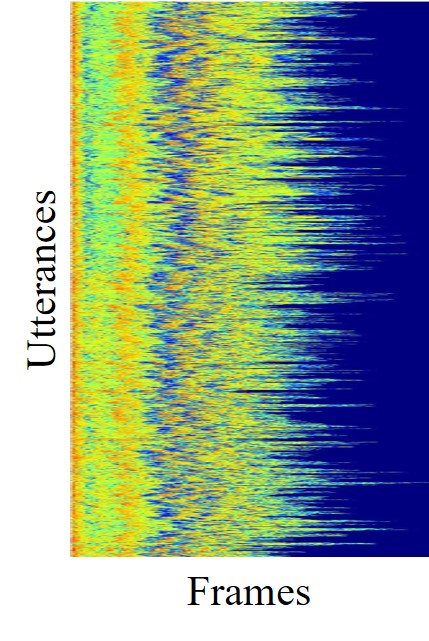}
		\caption{}
		\label{att_s_plot}
	\end{subfigure}
	\hspace{0.5cm}
	\begin{subfigure}[b]{0.2\textwidth}
		\centering
		\includegraphics[width=\linewidth]{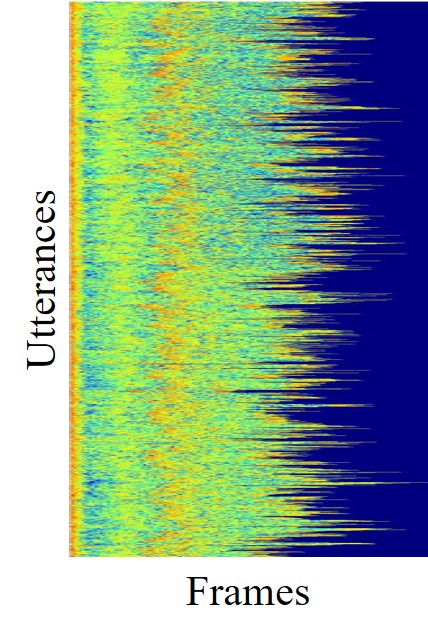}
		\caption{}
		\label{att_c_plot}
	\end{subfigure}
	\caption{Attention weights of \textit{d-vector (JFE)} for utterances speaking the sentence ``only lawyers love millionaires". (a) Attention weights for the speaker embedding vector. (b) Attention weights for the channel embedding vector.}
	\label{att_plot}
\end{figure}

In order to check if the training scheme of the proposed JFE system achieves our objective (i.e. maximizing the speaker discriminability and channel uncertainty in $\omega_{spkr}$), we analyzed the training loss described in (\ref{spkr_spkr_jvec})-(\ref{dev_spkr_jvec}) of the d-vector-based JFE system.
As shown in Fig. \ref{train_cost}, due to the large difference in the unique number of speakers and devices (i.e. 194 speakers and 6 devices), the initial values for  $\mathbf{L}_{s-s,CE}$ and $\mathbf{L}_{s-c,E}$ were higher than $\mathbf{L}_{c-c,CE}$ and $\mathbf{L}_{c-s,E}$.
The cross-entropy losses (i.e. $\mathbf{L}_{s-s,CE}$ and $\mathbf{L}_{c-c,CE}$) decreased quickly toward 0 when the training iteration increases.
On the other hand, the entropy losses (i.e. $\mathbf{L}_{s-c,E}$ and $\mathbf{L}_{c-s,E}$) stayed near at their initial values throughout the training.
This indicates that the proposed training scheme increases the discriminability of the speaker and channel embeddings on their main tasks while keeping their uncertainty on the subtasks high as expected.

In Fig. \ref{tsne_jvec}, the t-SNE plots \cite{tsne} of the speaker and channel embedding vectors of 10 speakers and 3 devices are shown.
As can be seen in Figs. \ref{speaker_plot_speaker_emb} and \ref{device_plot_speaker_emb}, the speaker embedding vectors ${\omega_{spkr}}$ were well separated between different speakers but were highly overlapped when it comes to different devices.
Meanwhile, as shown in Figs. \ref{speaker_plot_device_emb} and \ref{device_plot_device_emb}, the channel embedding vectors $\omega_{chan}$ were separately distributed in terms of the device, while they were inseparable in terms of speakers.
This confirms that the embedding vectors extracted from the proposed JFE system are discriminative on their main tasks, but are invariant with respect to their subtasks.

Moreover, in Fig. \ref{att_plot}, the attention weights for the utterance speaking the sentence ``only lawyers love millionaires" (i.e. $1^{st}$ sentence of the RSR2015 Part1 dataset) are shown.
It is interesting to see that the difference between speaker attention weights $\alpha_{spkr}$ across the frames were quite dramatic, which indicates that $\alpha_{spkr}$ are likely to attend to certain frames.
On the other hand, the channel attention weights $\alpha_{chan}$ were relatively consistent across all frames.
These results strongly support our assumption that the frames with high speaker-dependent information are concentrated on specific frames while channel-dependent information is similar across the speech segment.

\subsubsection{Comparison between the joint factor embedding scheme and conventional disentanglement methods}

\begin{table}[b]
	\renewcommand{\arraystretch}{1.3}
	\caption{EER (\%) comparison between the speaker embedding vectors extracted from the proposed joint factor embedding and the other embedding techniques.}
	\label{experiments}
	\centering
	\begin{tabular}{c|c|c}
		\hline
		             & Objective         & EER [\%]  \\ \hline
		\multirow{4}{*}{d-vector} & \textit{Softmax}           & 10.72 \\ 
		& \textit{Gradient reversal} & 10.37 \\ 
		& \textit{Anti-loss}         & 10.47 \\ 
		& \textit{\textbf{JFE (proposed)}}               & \textbf{8.43}  \\ \hline
		\multirow{4}{*}{x-vector} & \textit{Softmax}           & 2.26  \\ 
		& \textit{Gradient reversal} & 5.87  \\ 
		& \textit{Anti-loss}         & 1.46  \\ 
		& \textit{\textbf{JFE (proposed)}}               & \textbf{1.07}  \\ \hline
	\end{tabular}
\end{table}

\begin{table}[b]
	\renewcommand{\arraystretch}{1.3}
	\caption{Gender-dependent EER (\%) comparison between the speaker embedding vectors extracted from the x-vector-based embedding systems and the state-of-the-art i-vector-based systems.}
	\label{experiments_hmm}
	\centering
\begin{tabular}{c|c|c}
	\hline
	\multirow{2}{*}{Methods}                     & \multicolumn{2}{c|}{EER [\%]}      \\ \cline{2-3} 
	& Male          & Female        \\ \hline
	\textit{x-vector (Softmax)}                  & 2.09          & 2.48          \\ \hline
	\textit{DNN i-vectors} \cite{univector}                       & 1.70          & 2.69          \\
	\textit{Uncertainty normalized HMM/i-vector} \cite{univector} & 1.52          & 1.77          \\ \hline
	\textit{x-vector (GRL)}                      & 3.75          & 4.17          \\
	\textit{x-vector (Anti-loss)}                & 1.25          & 1.66          \\ \hline
	\textit{\textbf{x-vector (JFE)}}             & \textbf{0.82} & \textbf{1.29} \\ \hline
\end{tabular}
\end{table}

In this experiment, we compared the embedding vectors obtained from the proposed joint factor embedding scheme, with those obtained from the conventional disentanglement techniques discussed in Section \ref{d-vector}.
The experimented training strategies are as follows:
\begin{itemize}
	\item {\it{Softmax}}: embedding extracted from an embedding network trained with softmax objective in (\ref{softmax_dvec}),
	\item {\it{Gradient reversal}}: embedding extracted from an embedding network trained with gradient reversal strategy as described in (\ref{grl_1}) where $\lambda$ was set to be 0 in the beginning and linearly increased every iteration, reaching 1 at the end of the training as in \cite{asr_grl1},
	\item {\it{Anti-loss}}: embedding extracted from an embedding network trained with anti-loss as described in (\ref{antiloss_device}) using the same adversarial training strategy described in \cite{antiloss},
	\item {\it{JFE (proposed)}}: speaker embedding extracted from the proposed JFE system trained with the discriminative loss functions in (\ref{spkr_spkr_jvec}) and (\ref{dev_dev_jvec}) and both the entropy-based as shown in (\ref{spkr_dev_jvec}) and (\ref{dev_spkr_jvec}) and the negative MAPC-based disentanglement losses in (\ref{mapc}).
\end{itemize}

Table \ref{experiments} show the performance of the d-vector and x-vector-based systems trained with the methods described above.
Generally, the \textit{Anti-loss} disentanglement strategy has shown performance enhancement, achieving a relative improvement of 35.39\% in terms of EER in the d-vector-based experiment.
On the other hand, \textit{Gradient reversal} method, showed only slightly improved or worse performance over \textit{softmax}.
Meanwhile, the speaker embedding extracted from the proposed JFE scheme yielded the best performance in all architectures (i.e., d-vector and x-vector), achieving a relative improvement of 18.39\% in EER compared to that of \textit{d-vector (softmax)}.
This indicates that the proposed JFE system is capable of disentangling complicated corruptions (i.e. corruption via channel) introduced by different recording devices.

In addition, Table \ref{experiments_hmm} show the performance comparison between the state-of-the-art embedding techniques for random digit strings speaker verification (i.e., \textit{DNN i-vectors} and \textit{Uncertainty normalized HMM/i-vector}) \cite{univector} and the x-vector-based embedding network trained with the proposed JFE scheme.
As shown in the results, \textit{Uncertainty normalized HMM/i-vector} performs better than the \textit{x-vector (softmax)} by a large margin.
This is mainly attributed to the fact that the \textit{Uncertainty normalized HMM/i-vector} is trained to model the within-digit variability and scored with prior knowledge on the set of digits being uttered within the test set.
Therefore it is not surprising that the \textit{x-vector (softmax)} performs worse than the HMM/i-vector system, since it is trained and evaluated with no information on the context.
However, despite the innate disadvantage of the x-vector framework in random digits strings speaker verification, the proposed \textit{x-vector (JFE)} outperformed the \textit{Uncertainty normalized HMM/i-vector} with an relative improvement of 46.05\% in terms of male trial EER.

\subsubsection{Device disentanglement in domain-mismatch scenario}

\begin{table}[b]
	\renewcommand{\arraystretch}{1.3}
	\caption{EER (\%) comparison between the speaker embedding vectors extracted from the proposed joint factor embedding and the conventional x-vector framework evaluated on the VoxCeleb1 evaluation set.}
	\label{crossdomain}
	\centering
	\begin{tabular}{l|c|c}
		\hline
		Objective &Training data &  EER [\%] \\
		\hline
		\hline
		
		\multirow{2}{*}{\textit{x-vector (softmax)}} & VoxCeleb1 &   11.6\\
		
		& RSR2015 & 22.6 \\
		\hline
		
		\textit{\textbf{x-vector (JFE)}} & RSR2015 & 19.9
		
	\end{tabular}
\end{table}

In this experiment, we compared the performance of the conventional x-vector and the proposed JFE system in a cross-domain text-independent speaker verification scenario.
More specifically, both embedding systems were trained using the entire RSR2015 dataset and evaluated on the VoxCeleb1 evaluation subset, which is a dataset collected from Youtube videos recorded from a wide variety of channel and environmental conditions (e.g., videos shot on hand-held devices, interviews from red carpets).

As depicted in Table \ref{crossdomain}, the embeddings extracted from systems trained with RSR2015 showed severe performance degradation.
Such degradation was likely caused by the vast variety of channel and environmental conditions within the VoxCeleb1, which are known to cause high within-speaker variability of the extracted speaker embedding vectors.
Although the RSR2015 dataset is recorded from multiple different devices, the number of recording devices is limited (i.e. 6 devices) and the speech samples are relatively noise-free since they were recorded in an office environment \cite{rsr2015, rsr2015_2, rsr_protocol}.
Therefore training the embedding system using only the RSR2015 dataset may be insufficient to tackle the challenging condition of the VoxCeleb1 evaluation set.
Hence the \textit{x-vector} system trained only for speaker discrimination using RSR2015 showed a relative decrement of 94.83\% in terms of EER compared to the network trained with the VoxCeleb1 training set.
On the other hand, the degredation of the \textit{JFE} system trained to disentangle the device factor from the speaker embedding was 71.55\%, which outperformed the \textit{x-vector} trained with the same dataset with a relative improvement of 11.95\%.
This indicates that even in a domain-mismatch scenario, the proposed \textit{JFE} is able to alleviate the performance degradation caused by recording device variability.

\subsection{Emotion disentanglement}
Emotion variability can cause severe performance degradation in speaker recognition \cite{emo}, but emotion disentanglement has not been investigated as much as other nuisance attributes, such as noise or channel distortion.
This may be due to the challenging nature of emotion disentanglement since unlike noise or channel, emotional variability is caused by the speaker's vocal tract, which also creates speaker variability.
In this subsection, we apply the proposed JFE framework for disentangling the variability induced by the speaker's emotional state.

\subsubsection{Dataset}
In order to evaluate the performance of the proposed technique for emotion disentanglement, a set of experiments were conducted based on the VoxCeleb1 dataset \cite{vox1} and the emotion labels provided by the EmoVoxCeleb teacher system \cite{emovox} \footnote[1]{The emotion labels provided by the EmoVoxCeleb teacher system can be downloaded from here: http://www.robots.ox.ac.uk/~vgg/research/cross-modal-emotions/.}.
For training the embedding networks, we used the \textit{development} subset of the VoxCeleb1 dataset, consisting of 148,642 utterances collected from 1,211 speakers.
According to the emotion labels in EmoVoxCeleb, total 8 emotions are observed in the VoxCeleb1 dataset (i.e., neutral, happy, surprise, sad, angry, disgust, fear, contempt).

The evaluation was performed according to the original VoxCeleb1 trial list, which consists of 4,874 utterances spoken by 40 speakers.
The duration of the trial utterances was between 3.97 seconds and 69.05 seconds.

\subsubsection{Experimental Setup}
The acoustic features used in the experiments were 30-dimensional MFCCs extracted at every 10 ms, using a 20 ms Hamming window.
The embedding networks were trained with segments consisting of 250 frames, using the ADAM optimization technique.

For the baseline x-vector framework and joint factor embedding system, 5 TDNN layers were used as the frame-level network according to the Kaldi x-vector recipe \cite{xvec}.
The TDNN outputs are aggregated as described in (\ref{attention_1}), and fed into the utterance-level classification network (i.e. speaker and emotion identifier).
Each utterance-level classification network consisted of two 512-dimensional LeakyReLU hidden layers and a softmax output layer where the output size corresponds to the number of speakers or emotions within the training set.
All the experimented networks were trained with learning rate 0.001 and batch size 256 for 74,321 iterations.
Cosine similarity was used for computing the verification scores in the experiments.

\subsubsection{Comparison between the joint factor embedding scheme and conventional embedding techniques}

\begin{table*}[b]
	\renewcommand{\arraystretch}{1.3}
	\caption{EER (\%) comparison between the speaker embedding vectors extracted from the proposed joint factor embedding and the conventional methods. In the Data Augmentation column, X indicates embedding network trained with no augmentation and O indicates network trained with augmented training set.}
	\label{vox_performance}
	\centering
	\begin{tabular}{c|c|c|c}
		\hline
		\multicolumn{1}{c|}{Methods}                         & \multicolumn{1}{c|}{Scoring} & Data augmentation & EER {[}\%{]} \\ \hline
		\textit{i-vector} \cite{xvec_perf}                                   & PLDA                         & X                 & 8.8          \\ \hline
		\textit{VGG} \cite{xvec_perf}                                        & Cosine similarity            & X                 & 7.8          \\ \hline
		\textit{Generalized end-to-end} \cite{ge2e_perf}                     & Cosine similarity            & X                 & 10.7         \\ \hline
		\textit{All-speaker hard negative mining end-to-end} \cite{ge2e_perf} & Cosine similarity            & X                 & 5.6          \\ \hline
		\multirow{3}{*}{\textit{x-vector (softmax)} \cite{xvec_perf}}         & Cosine similarity            & X                 & 11.3         \\ \cline{2-4} 
		& \multirow{2}{*}{PLDA}        & X                 & 7.1          \\ \cline{3-4} 
		&                              & O                 & 6.0          \\ \hline
		\textit{x-vector (our implementation)}                                        & PLDA            & O                 & 4.9          \\ \hline
		\multirow{3}{*}{\textit{CNN-embedding} \cite{xvec_perf}}              & Cosine similarity            & X                 & 7.3          \\ \cline{2-4} 
		& \multirow{2}{*}{PLDA}        & X                 & 5.9          \\ \cline{3-4} 
		&                              & O                 & 5.3          \\ \hline
		\multirow{3}{*}{\textit{\textbf{x-vector (JFE)}}}    & Cosine similarity            & X                 & \textbf{6.8} \\ \cline{2-4} 
		& \multirow{2}{*}{PLDA}        & X                 & \textbf{5.4} \\ \cline{3-4} 
		&                              & O                 & \textbf{4.4} \\ \hline
	\end{tabular}
\end{table*}

\begin{figure*}
	\centering
	\includegraphics[width=0.9\linewidth]{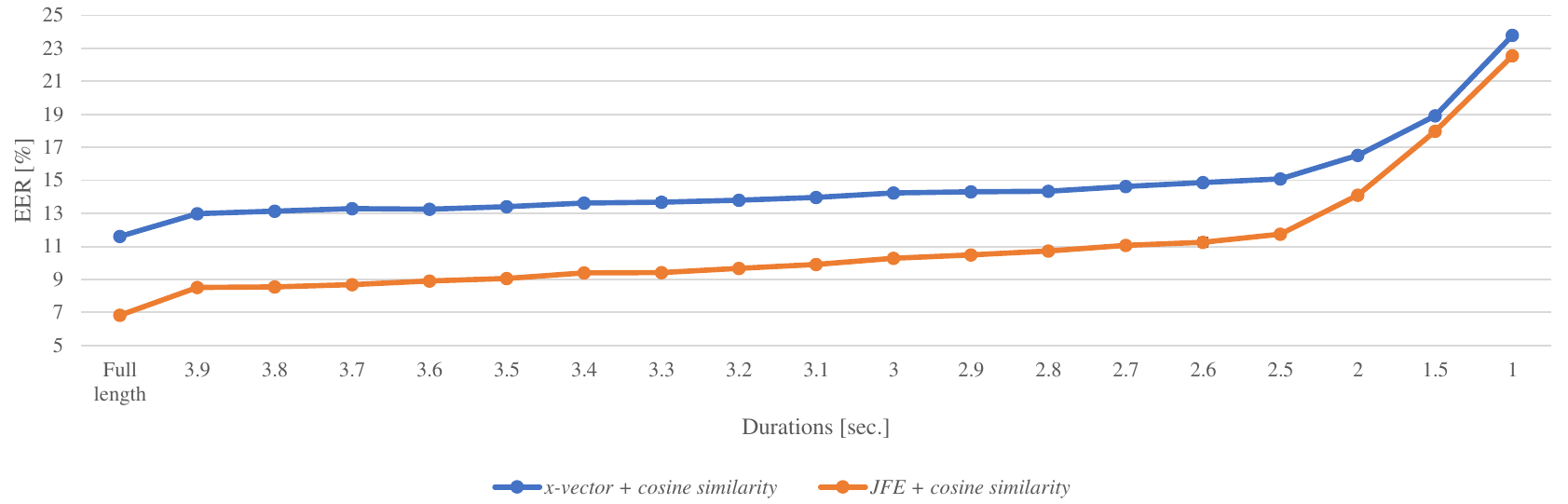}
	\caption{EER performance of the proposed joint factor embedding scheme and conventional x-vector on different duration utterances.}
	\label{duration_chart}
\end{figure*}

In this experiment, we compare the embedding vectors obtained from the proposed joint factor embedding scheme and the conventional x-vector framework along with techniques reported in recent studies including VGG-M, ResNet-34 and end-to-end verification systems \cite{xvec_perf, ge2e_perf}.
The experimented methods are as follows:
\begin{itemize}
	\item \textit{i-vector} \cite{xvec_perf}: the i-vector performance reported in \cite{xvec_perf},
	\item \textit{VGG} \cite{xvec_perf}: the performance of the embedding extracted from VGG-M, which is a CNN architecture known to perform well on image and speaker classification, reported in \cite{xvec_perf},
	\item \textit{Generalized end-to-end} \cite{ge2e_perf}: the performance of the ResNet-34-based end-to-end speaker verification system trained with the generalized end-to-end loss (\ref{e2e_dvec_2}) reported in \cite{xvec_perf},
	\item \textit{All-speaker hard negative mining end-to-end} \cite{ge2e_perf}: the performance of the ResNet-34-based end-to-end speaker verification system trained with the all-speaker hard negative mining loss, which is a modified version of the softmax loss for robust verification, reported in \cite{xvec_perf},
	\item \textit{x-vector (softmax)} \cite{xvec_perf}: the x-vector performance reported in \cite{xvec_perf},
	\item \textit{x-vector (our implementation)}: the performance of our implementation of \textit{x-vector (softmax)},
	\item \textit{CNN-embedding} \cite{xvec_perf}: the performance of the embedding extracted from a CNN-based architecture reported in \cite{xvec_perf},
	\item \textit{x-vector (JFE)}: the performance of the speaker embedding extracted from the proposed JFE system trained to disentangle the emotional factor using loss functions (\ref{spkr_spkr_jvec})--(\ref{mapc}).
\end{itemize}

As shown in Table \ref{vox_performance}, the proposed \textit{JFE} outperformed the conventional methods with both cosine similarity and PLDA backends.
Especially when using PLDA as backend, the \textit{JFE} achieved a relative improvement of 8.16\% compared to the \textit{x-vector (our implementation)} in terms of EER.
Moreover, training the \textit{JFE} with augmented training data described in \cite{xvec_perf} (i.e., noise and reverberation augmentation) further improved the performance.
The results demonstrate that although the proposed \textit{JFE} is composed of a simple x-vector-like network, it can provide embedding with higher speaker discriminative information than the systems with more complicated architecture.

In addition, we evaluated the conventional x-vector framework and the proposed joint factor embedding scheme on short duration speech samples.
Each evaluation was done using randomly truncated trial utterances and the average EERs computed over three evaluations for each duration group are depicted in Fig. \ref{duration_chart}.
As shown in the results, both the performance of the joint factor embedding framework and the conventional x-vector were degraded as the duration decreased.
This may be due to the lack of phonetically informative frames since a critical amount of speaker relevant information is contained in the phonetic characteristics \cite{duration}.
However, the emotion disentangled speaker embedding obtained by the proposed \textit{JFE} outperformed the conventional \textit{x-vector} even with short duration speech segments.

\section{Conclusion} \label{conclusion}

In this paper, a novel approach for extracting an embedding vector robust to variability caused by nuisance attributes for speaker verification is proposed.
In order to disentangle the nuisance variability from the speaker embedding vector, we introduce a JFE scheme where two types of embedding vectors are extracted, each dependent solely on the speaker or nuisance attribute, respectively.
The proposed JFE network is trained simultaneously with the speaker and nuisance attribute classification networks where the speaker and nuisance embedding vectors are optimized to have good discriminability for their main task while having high uncertainty on their subtask.

To evaluate the performance of the embedding vector extracted from the proposed system in a realistic scenario, we conducted a set of speaker verification experiments using the RSR2015 dataset, which is composed of utterances recorded using multiple different hand-held devices, and VoxCeleb1 dataset, which is composed of various emotional speech utterances.
From the results, it is shown that the proposed JFE scheme is capable of obtaining speaker embedding vectors with high speaker discriminability while showing robustness to channel and emotional variability.
Moreover, we observed that the proposed embedding vector performs better than the conventional embedding technique with short duration speech segments.

Although the proposed technique showed great improvement over the conventional methods, since the proposed JFE is trained in a fully supervised manner, it requires labels for not only the speakers but also the nuisance attributes.
Thus in our future study, we will expand the JFE technique to disentangle the non-speaker variability without the supervision of nuisance attribute labels.
Moreover, we will improve the disentanglement performance by using more sophisticated methods for reducing the mutual information between the speaker and nuisance embedding vectors, rather than using a simple MAPC regularization.

\section*{Acknowledgment}
This work was supported by the BK21 Plus program of the Creative Research Engineer Development for IT, Seoul National University in 2020.
This research was supported and funded by the Korean National Police Agency. [Project Name: Real-time speaker recognition via voiceprint analysis / Project Number: PA-J000001-2017-101]

\begin{IEEEbiography}{Woo Hyun Kang}
	was born in Seoul, Korea, in 1990. He received the B.S. degree in electronics engineering from Kookmin University, Seoul, Korea, in 2014. He is currently pursuing the Ph.D. degree in electrical engineering and computer science at Seoul National University (SNU), Seoul, Korea. 
	
	His research interests include speaker recognition, machine learning, and signal processing.
\end{IEEEbiography}

\begin{IEEEbiography}{Sung Hwan Mun}
	was born in Incheon, Korea, in 1993. He received the B.S. degree in electronics engineering from Inha University, Incheon, Korea, in 2019. He is currently pursuing the Ph.D. degree in electrical engineering and computer science at Seoul National University (SNU), Seoul, Korea.
	
	His research interests include speaker recognition, machine learning, and signal processing.
\end{IEEEbiography}

\begin{IEEEbiography}{Min Hyun Han}
	was born in Seoul, Korea, in 1992. He received the B.S. degree in electrical \& electronic engineering from Yonsei University, Seoul, Korea, in 2018. He is currently pursuing the Ph.D. degree in electrical engineering and computer science at Seoul National University (SNU), Seoul, Korea.
	
	His research interests include speaker recognition, machine learning, and signal processing.
\end{IEEEbiography}

\begin{IEEEbiography}{Nam Soo Kim}
	received the B.S. degree in electronics engineering from Seoul National University (SNU), Seoul, Korea, in 1988 and the M.S. and Ph.D. degrees in electrical engineering from Korea Advanced Institute of Science and Technology in 1990 and 1994, respectively.
	
	From 1994 to 1998, he was with Samsung Advanced Institute of Technology as a Senior Member of Technical Staff. Since 1998, he has been with the School of Electrical Engineering, SNU, where he is currently a Professor. His research area includes speech signal processing, speech recognition, speech/audio coding, speech synthesis, adaptive signal processing, machine learning, and mobile communication.
\end{IEEEbiography}

\EOD

\end{document}